\documentstyle[12pt,psfig]{article}
\normalsize
\begin{document}

\begin{titlepage}
\begin{center}
{\bf \large 
Cascade-exciton model analysis \\
of proton spallation  from 10 MeV to 5 GeV\\
} 
\vspace*{1.2cm}
{ S.~G.~Mashnik$^{*}$ and A.~J.~Sierk}\\
{\em T-2, Theoretical Division, Los Alamos National Laboratory,
Los Alamos, NM 87545}\\
\vspace*{.3cm}
{O.~Bersillon}\\ 
{\em CEA, Centre d'Etude de Bruy\`eres-le-Ch\^atel, 91680,
Bruy\`eres-le-Ch\^atel, France}\\
\vspace*{.25cm}
{and}\\
\vspace*{.25cm}
{ T.~Gabriel}\\
{\em Oak Ridge National Laboratory, Oak Ridge, Tennessee 37831}\\
\vspace{2.0cm}
Abstract\\
\end{center}
\vspace*{10pt}
\noindent{
We have used an extended version of the Cascade-Exciton Model (CEM)
to analyze more than 600 excitation functions for proton induced
reactions on 19 targets ranging from $^{12}$C to $^{197}$Au, for
incident energies ranging from 10~MeV to 5~GeV.  We have compared
the calculations to available data, to calculations using 
approximately two dozen other models, and to predictions of several
phenomenological systematics.  We present here our
conclusions concerning the relative roles of different reaction
mechanisms in the production of specific final nuclides. We comment 
on the strengths and weaknesses of the CEM and suggest possible
further improvements to the CEM and to other models.
}

\vspace{175pt}
\hspace*{-10mm} 
\rule[5mm]{70mm}{0.01mm}
\hspace*{-71mm} 
$^{*}$On leave of absence from Bogoliubov Laboratory of Theoretical
Physics, \\
\hspace*{1mm} 
Joint Institute for Nuclear Research, Dubna, Russia

\end{titlepage}

\newpage
 
Precise nuclear data on excitation functions for reactions
induced by nucleons in the energy range up to several GeV are of
great importance both for fundamental nuclear physics and for
many applications.  Such data are necessary to understand the
mechanisms of nuclear reactions, to study the change of properties
of nuclei with increasing excitation energy, and to study the
effects of nuclear matter on the properties of hadrons and their
interactions. Excitation functions are more sensitive to the 
detailed mechanisms
of nuclear reactions than are double differential cross sections
of emitted particles or their integrals over energy and/or angles.
Therefore, excitation functions are a convenient tool to test
models of nuclear reactions.

Second, and perhaps more important today, expanded nuclear data bases 
in this intermediate energy range are required for several 
important applications.
Recently, one of the most challenging problems 
requiring reliable nuclear data files is
Accelerator-Driven Transmutation Technology (ADTT)
for elimination of nuclear waste \cite{c1}.
The problems of Accelerator Transmutation of Waste (ATW) are
closely connected
with Accelerator-Based Conversion (ABC)~\cite{c2} aimed to
complete the destruction of weapons plutonium, and with 
Accelerator-Driven Energy Production (ADEP) 
\cite{c3} which
proposes to derive fission energy from thorium with concurrent
destruction of the long-lived waste and without the
production of weapons-usable material,
though substantial differences among these systems do exist~\cite{c2}.
Precise nuclear data are needed for solving problems of radiation 
damage to microelectronic devices~\cite{sosnin93} and not only of
radiation protection of cosmonauts and aviators or workers at 
nuclear installations, but also to estimate the radiological impact of
radionuclides such as $^{39}$Ar arising from the operation of fusion reactors
or high-energy accelerators and the population dose from such radionuclides
retained in the atmosphere so as to avoid possible problems of radiation
health effects for the whole population (see, e.g.~\cite{ki92}).
Another important new application which requires large nuclear data 
libraries at energies up to several hundreds of MeV is the radiation
transport simulation of cancer radiotherapy used for selecting the 
optimal dose in clinical treatment planning systems~\cite{white94}.
Many excitation functions are needed for the optimization 
of commercial production
of radioisotopes widely used in different branches of nuclear medicine
\cite{c7}, mining and industry~\cite{edmonts86}. Also,
residual product nuclide yields in thin targets irradiated by medium-
and high-energy projectiles are extensively used in cosmochemistry and
cosmophysics, e.g. to interpret the production of cosmogenic nuclides in 
meteorites by primary galactic particles~\cite{michel96},
etc.

Because of the impracticality of measuring all cross sections
important to the processes of pragmatic interest, it is important to 
try to develop 
reliable models to predict cross sections which have not
been or cannot be measured.  In order to carefully evaluate the
strengths and weaknesses of one such model, we have undertaken a
careful comparison of an extended Cascade-Exciton Model~\cite{cem}
as realized in the CEM95 code with both experimental data
on excitation functions for proton-induced reactions and with many
other model calculations.   We have studied the dependence of our
results on the physics incorporated in the code, on the values of
input parameters, on the incorporation of the isotopic
composition of actual experimental targets, and on the proper
modeling of independent and cumulative yields.

We have performed detailed analyses of more than 600 excitation 
functions for interactions
of protons with energies from 10~MeV to 5~GeV with nuclei of
$^{12}$C, $^{14}$N, $^{16}$O, $^{27}$Al, $^{31}$P, $^{40}$Ca, $^{54}$Fe,
$^{56}$Fe, $^{57}$Fe, $^{58}$Fe, $^{nat}$Fe, $^{59}$Co, $^{90}$Zr,
$^{91}$Zr, $^{92}$Zr, $^{94}$Zr, $^{96}$Zr, $^{nat}$Zr, and  $^{197}$Au.
We have compared our results with all reliable experimental data 
available to us and with predictions of other models realized in 
several codes: 
ALICE LIVERMORE 87~\cite{alicelivermore87}, 
HETC/KFA-2~\cite{hetckfa2},
ALICE91~\cite{alice91},
LAHET~\cite{lahet},
ALICE-F~\cite{alicef},
NUCLEUS~\cite{nucleus},
MCEXCITON~\cite{mcexciton},
ALICE82~\cite{alice82},
DISCA2~\cite{disca2},
CASCADE~\cite{cascade},
HETC~\cite{hetc},
INUCL~\cite{inucl},
ALICE75~\cite{alice75},
ALICE LIVERMORE 82~\cite{alicelivermore82},
ALICE 87 MOD~\cite{alice87mod},
PEQAQ2~\cite{peqaq2},
ALICE92~\cite{alice92},
CEM92M~\cite{cemphys},
with the Milan version of the exciton model of
nuclear reactions with preformed $\alpha$-clusters in nuclei~\cite{ga77},
and with calculations using phenomenological 
systematics from Refs.~\cite{ru66}--\cite{ko93}.
A comparison of many of our results with predictions of several other
codes may be found in a recent NEA/OECD document~\cite{mn97}.
A comparison of the 
yields of residual product nuclei in $^{209}$Bi thin targets
irradiated
by 130 MeV and 1.5 GeV protons simulated by CEM95 with the recent
measurements by Titarenko et al. \cite{titarenko97} and with results
obtained with the codes 
HETC~\cite{hetc},
GNASH~\cite{gnash},
LAHET~\cite{lahet},
INUCL~\cite{inucl},
CASCADE~\cite{cascade}, 
and
ALICE96~\cite{alice96}
may be found above in this issue in the previous paper \cite{titarenko97}. 

A detailed report of the study~\cite{report97},
containing 179 pages, 103 figures, and 243 references to 308 original
books, journal articles, preprints, theses, and conference contributions
is available on the World Wide Web as a compressed PostScript file,
or in hard copy from either of the first two authors.  Here we will
present only our main conclusions from the study.

Our analyses have shown that several different mechanisms 
participate in the production of most final nuclides.
Their relative roles change significantly with the changing atomic mass
of the targets, with increasing
incident energy, and are different for different final nuclides. 
The main nuclide production mechanism in the spallation region is the
successive emission of several nucleons, while emission of complex
particles is important (and may be even the only
mechanism for production of a given isotope in a limited range
of incident energy) only at low incident energies, near the corresponding 
thresholds, while with increasing energy its relative role 
decreases quickly.

For medium and especially for heavy targets,
the contribution from radioactive precursors to the
measured yields of many nuclides is very important.
The cumulative yields of some nuclides are up to two orders of magnitude
higher than the independent ones. Therefore, for heavy targets, 
especially careful calculations of cumulative yields and their
comparisons with the measured data are needed.

Our analyses have shown that nuclear structure effects are
very important in production of some nuclides and manifest
themselves strongly even at an incident energy of 5 GeV. 
Therefore, reliable and well fitted models of shell and pairing
corrections, level density parameters, and especially of nuclear
masses and consequent binding energies and $Q$-values have to
be used in calculations. 

The extended version of the cascade-exciton model
realized in the code CEM95 describes satisfactorily
with a fixed set of input parameters the shapes and  
absolute values of the majority of measured excitation functions
for production of nuclides in the spallation region and for
the emission of secondary nucleons and complex particles. 
We feel that the yields of both nuclides in the spallation
region and secondary particles of $A < 4$ predicted by CEM95 are at 
least as reliable, and in many cases more so, than those of the
other models and phenomenological systematics mentioned above.

For target nuclei from $^{27}$Al to $^{197}$Au, CEM95 describes
the majority of experimental excitation functions 
in the spallation region to within a factor of 2.  For targets 
lighter than $^{27}$Al, the agreement with experimental data
is worse, and the CEM, like the majority of other models, 
has to be improved to be able to describe
excitation functions from light targets.
Because CEM95 does not contain a special mechanism for fragmentation,
because it
underestimates production of $^4$He, and does not include a model
of fission fragment production, it cannot reliably predict
nuclide yields in the mass and energy regions where these
processes are dominant. These mechanisms of nuclear reactions
will need to be incorporated into the CEM.

In rare cases, in the same spallation region where it is usually
reliable, CEM95 underestimates or overestimates some individual
measured excitation functions, sometimes up to an order of magnitude.
This is mainly a result of the poor nuclear mass and binding energy values
used in CEM95. 

We conclude that the extended version of the
cascade-exciton model realized in the code CEM95 is
suitable for a rough evaluation of excitation functions in the
spallation region. But for a better
description of the measured yields in this region and
for an extension of the range of its applicability into the
fission and fragmentation regions, it should be developed further.
Among improvements of the CEM which are of highest priority
we consider the following:

\begin{itemize}
\item incorporation of recent experimental nuclear mass tables, and new 
   reliable theoretical mass formulas for unmeasured nuclides,
\item development and incorporation of an appropriate model of 
   high-energy fission,
\item modeling the emission of gammas competing with the evaporation of 
   particles at the compound stage,
\item treating more accurately $\alpha$-emission at the
   preequilibrium stage,
\item incorporation of a model for fragmentation of medium and heavy
   nuclei, and the Fermi breakup model for highly excited light nuclei,
\item modeling the evaporation of fragments with $A > 4$ from not 
   too light excited nuclei (incorporation of such processes at the
   preequilibrium stage may also be important),
\item modeling the coalescence of light fragments from
   fast emitted particles,
\item improvement of the approximations for inverse cross sections, and
\item use of new, more precise experimental data for the cross sections
   of elementary interactions at the cascade stage.
\end{itemize}
   
Such a development and improvement of the CEM is possible, and 
work in this direction is already in progress.
We hope that a proper incorporation of the above improvements in the
code will not destroy the present wholeness of CEM95 
and its good predictive power for the spectra of secondary particles.

There are a number of other possible and desirable improvements of the
CEM discussed in the complete report~\cite{report97}, which are
justified from a physical point of view.
Unfortunately, the inclusion of separate refinements in nuclear models 
used in INC calculations does
not always lead to improved agreement with experimental data.

The problems discussed above are typical not only of the CEM,
but also for all other similar models and codes, where they are also not
solved yet. Excitation functions are a very ``difficult" characteristic
of nuclear reactions as they involve together the different and complicated
physics processes of spallation, evaporation, fission, and fragmentation 
of nuclei. A lot of work is still necessary to be done by
theorists and code developers before a reliable complex of codes 
able to satisfactorily predict arbitrary excitation functions 
in a wide range of  
incident energies/projectiles/targets/final nuclides
will be available. At present, we are still very far from the completion 
of this difficult task.

In the meantime, to evaluate excitation functions needed for
science and applications, it is necessary to 
use and analyze together the available experimental data, and for each
region of incident energies/projectiles/targets/final nuclides,
the predictions of phenomenological systematics and
the results of calculations with the most reliable codes.
Our present study has shown that for proton-induced
reactions in the spallation region, not too low incident
energies and not too light targets, CEM95 is such a reliable code.\\

{\large \bf Acknowledgements} \\

It is a pleasure to acknowledge M.~Blann, V.~P.~Eismont, L.~M.~Krizhansky,
R.~Michel, P.~Nagel, H.~S.~Plendl and Yu.~E.~Titarenko
for useful discussions on spallation
physics which have stimulated us to carry out this research.  
One of the authors (S.~G.~M.) thanks CEA, Bruy\`eres-le-Ch\^atel
and ORNL, Oak Ridge for kind hospitality and excellent
conditions during his work in February--August 1996 at
Bruy\`eres-le-Ch\^atel and in July--August 1995 at Oak Ridge,
where most of this study was done.
He is also grateful to M.~B.~Chadwick, R.~E.~MacFarlane, D.~G.~Madland,
P.~M\"oller, J.~R.~Nix, R.~E.~Prael, L.~Waters, and P.~G.~Young of
LANL for fruitful discussions and support.

This study was completed under the auspices of the U.S.
Department of Energy by the Los Alamos National Laboratory under
contract no.~W-7405-ENG-36.

\newpage

\end{document}